\documentclass[10pt]{IEEEtran}

\usepackage{amsmath}
\usepackage{mathtools}
\usepackage{graphicx}
\usepackage{enumerate}
\usepackage{url}
\usepackage{subfigure}
\usepackage{algorithm}
\usepackage{algpseudocode}
\usepackage{stfloats}
\usepackage{lineno}
\usepackage{hyperref}
\usepackage{bm}
\usepackage{bbm}
\usepackage{amssymb}
\usepackage{enumerate}
\usepackage{xcolor}
\usepackage{float}
\usepackage{cite}
\usepackage{tabularx}
\usepackage{tabu}
\usepackage{multicol}
\usepackage{multirow}
\usepackage{colortbl,booktabs,threeparttable}
\usepackage{dcolumn}
\usepackage{flushend}
\usepackage{soul}
\usepackage{ragged2e}

\graphicspath{{Figures/}}

\renewcommand{\vec}[1]{\bm #1}

\newcommand{\mat}[1]{\bm #1}

\newcommand{\vech}[1]{\hat{\bm #1}}

\renewcommand{\math}[1]{\hat{\bm #1}}

\renewcommand{\cal}[1]{\mathcal{#1}}

\newcommand{\E}{\mathbb{E}}
\newcommand{\D}{\mathbb{D}}
\renewcommand{\P}{\mathbb{P}}

\newcommand{\Ph}{\hat{\mathbb{P}}}

\newcommand{\Pb}{\bar{\mathbb{P}}}

\renewcommand{\Pr}{\operatorname{Pr}}

\renewcommand{\H}{\mathsf{H}}
\renewcommand{\st}{\text{s.t.}}

\newcommand{\defeq}{\coloneqq}

\newcommand{\tabincell}[2]{\begin{tabular}{@{}#1@{}}#2\end{tabular}}
\newcommand{\quotemark}[1]{``#1”}

\definecolor{hl-bg-color}{RGB}{255,255,215}
\sethlcolor{hl-bg-color} 
\definecolor{new-magenta}{RGB}{255,0,255}
\soulregister\cite7
\soulregister\citep7
\soulregister\citet7
\soulregister\ref7
\soulregister\eqref7
\ifdefined\HIGHLIGHT
    \newcommand{\red}[1]{{\color{red}{#1}}}
    \newcommand{\blue}[1]{{\color{blue}{#1}}}

\else
    \renewcommand{\hl}[1]{{#1}}
    \newcommand{\red}[1]{{#1}}
    \newcommand{\blue}[1]{{#1}}

\fi

\begin{document}
\newpage
\title{Uncertainty Awareness in Wireless Communications and Sensing}

\author{Shixiong Wang, 
        Wei Dai,
        Jianyong Sun,
        Zongben Xu,
        and Geoffrey Ye Li,~\IEEEmembership{Fellow,~IEEE} 
\thanks{S.  Wang, W. Dai, and G. Li are with the Department of Electrical and Electronic Engineering, Imperial College London, London SW7 2AZ, United Kingdom (E-mail: 
s.wang@u.nus.edu; 
wei.dai1@imperial.ac.uk; 
geoffrey.li@imperial.ac.uk); J. Sun and Z. Xu are with the School of Mathematics and Statistics, Xi’an Jiaotong University,
Xi’an 710049, China.
(E-mail: jy.sun@xjtu.edu.cn; zbxu@xjtu.edu.cn).
(\textit{Corresponding Author: S. Wang.})
}
\thanks{This work is supported by the UK Department for Science, Innovation
and Technology under the Future Open Networks Research Challenge project
TUDOR (Towards Ubiquitous 3D Open Resilient Network). 
}
}

\maketitle

\begin{abstract}
Wireless communications and sensing (WCS) establish the backbone of modern information exchange and environment perception. Typical applications range from mobile networks and the Internet of Things to radar and sensor grids. 
Despite transformative capabilities, wireless systems often face diverse uncertainties in design and operation, such as modeling errors due to incomplete physical knowledge, statistical errors arising from data scarcity, measurement errors caused by sensor imperfections, computational errors owing to resource limitation, and unpredictability of environmental evolution. Once ignored, these uncertainties can lead to severe outcomes, e.g., performance degradation, system untrustworthiness, inefficient resource utilization, and security vulnerabilities. As such, this article reviews mature and emerging architectural, computational, and operational countermeasures, encompassing uncertainty-aware designs of signals and systems (e.g., diversity, adaptivity, modularity), as well as uncertainty-aware modeling and computational frameworks (e.g., risk-informed optimization, robust signal processing, and trustworthy machine learning). Trade-offs to employ these methods, e.g., robustness vs optimality, are also highlighted. 
\end{abstract}


\section{Introduction} \label{sec:introdction}
\IEEEPARstart{T}{he} 20th and 21st centuries have witnessed the advent, development, and maturation of wireless systems that play key roles in communications and sensing. Representative examples include cellular networks, satellite systems, Bluetooth meshes, wireless fidelity (Wi-Fi) systems, the Internet of Things (IoT), autonomous swarms, radar, and sensor networks, to name a few. These wireless systems have greatly revolutionized human society by enabling the intelligent acquisition and exchange of information, for example, high-speed data transfer, ubiquitous connectivity, environmental understanding, medical microwave imaging, and remote healthcare. \blue{The incorporation of advanced analysis and processing techniques, such as artificial intelligence and machine learning, further pushes the boundaries of wireless communications and sensing}, enabling automated and high-quality data analytics, and supporting sensible and efficient decision-making \cite{wang2020thirty}.

Although significant functions and capabilities have been shown, wireless systems often face numerous modeling, experimental, computational, operational, and environmental uncertainties that threaten their trustworthiness, efficiency, and security. \blue{Here, uncertainties refer to discrepancies between our knowledge and the underlying truths.} These uncertainties can arise from diverse sources, for instance, channel modeling errors, sensor measurement errors, algorithmic truncation errors, network interferences and attacks, and environmental evolutions and fluctuations, respectively. In the design and operation of wireless systems, if uncertainties are ignored, the consequences can be severe in the sense of conspicuous performance degradation, system untrustworthiness, inefficient resource utilization, and security vulnerabilities. To be specific, for example, wireless signals inevitably undergo non-stationary random channel fading (e.g., time-varying Rayleigh or Rician), hardware imperfections (e.g., nonlinearities in power amplifiers, in-phase--quadrature imbalances, and phase noises), and interferences from other devices (e.g., co-channel, jamming, and network attacks), which can cause unpredictable variations in signal quality and uncontrollable errors in signal detection, estimation, and analysis, if these uncertainties are not sufficiently handled. For another example, when external environments evolve or fluctuate, the performance of employed signal processing and machine learning methods for communications and sensing may seriously degrade, if the adaptation or robustification of these methods is not addressed. Therefore, uncertainty awareness comprises a critical aspect of intelligent transmission and processing; see \cite[Fig.~10]{wang2024machine}.

To enhance and ensure the trustworthiness, efficiency, and security of wireless systems, great efforts have been made in the design and operation processes by both academia and industry. Aiming to illuminate the path to uncertainty-aware (UA) transmission and processing, this article reviews mature and emerging architectural, computational, and operational mitigation strategies in response to different types of uncertainties in wireless communications and sensing. 
We categorize the existing and developing UA treatments into two primary streams. 
The first stream works on reforms of UA wireless signals and systems, considering the following key factors:
\begin{itemize}
    \item  Redundancy, Diversity, and Margin; 
    \item  Feedback and Adaptivity; 
    \item  Anomaly Detection and Handling; 
    \item  Modularization; 
    \item  Decentralization; 
    \item  and Prediction and Prescription. 
\end{itemize}
\blue{These solutions emphasize modifying signal characteristics (e.g., structures, parameters), system architectures, and operational strategies to enhance uncertainty awareness.} The second stream focuses on enriching UA modeling and computational frameworks, emphasizing the following aspects:
\begin{itemize}
    \item  Uncertainty Quantification; 
    \item  UA Optimization; 
    \item  Adaptive and Robust Signal Processing; 
    \item  and Trustworthy Machine Learning. 
\end{itemize}
\blue{These solutions are particularly useful when a wireless communication or sensing problem can be formulated as an optimization, signal processing, or machine learning model.}

However, advanced system characteristics, such as adaptivity and robustness, always come with additional prices, e.g., higher design expenses, resource idling for overengineering, loss of performance optimality under ideal conditions, and extra computational burdens. Hence, in practice, balances among conflicting system features must be carefully planned. To this end, we also summarize potential trade-offs in UA wireless systems engineering.

This article is organized as follows. Section \ref{sec:sources} identifies diverse sources of uncertainties in wireless communications and sensing, and highlights the necessity of uncertainty awareness. Section \ref{sec:technical} outlines architectural, computational, and operational approaches to address these uncertainties. Trade-offs in UA wireless engineering are discussed in Section \ref{sec:prices}, while concrete examples of uncertainty awareness are given in Section \ref{sec:examples}. 
Conclusions in Section \ref{sec:conclusion} complete the article.

\section{Sources of Uncertainties and Necessity of Uncertainty Awareness}\label{sec:sources}
\blue{Uncertainties arise when our knowledge deviates from the underlying truths, for example,} discrepancies between modeling assumptions and physical mechanisms, between observations and actual values, between found local optimality and unknown global optimality, and between limited history data and population distribution. The more complex and dynamic a system, model, process, or population is, the more uncertainties may appear. \blue{Based on their origins}, uncertainties in wireless systems can be exemplified as follows, along with the potential consequences if not properly handled.

\textit{Channel Uncertainties}: Channel uncertainties can be caused by modeling errors (e.g., mismatched modeling assumptions), moving objects, environmental evolutions and fluctuations, and estimation errors due to data scarcity and limited computing resources. Ignoring channel uncertainties may lead to severe repercussions, such as dead zones in mobile networks, weak connectivity, reduced or unstable signal-to-noise ratio (SNR), inter-symbol interference, frequency offsets (i.e., Doppler), and inefficient resource utilization \blue{(e.g., high spectrum and power consumption with limited performance gains)}.

\textit{Noises, Interferences, and Attacks}: In wireless systems design and operation, thermal noises, quantization noises, phase noises, co-channel interference, adjacent-channel interference, inter-system (e.g., between Bluetooth and Zigbee) interference, self-interference (e.g., in full-duplex systems), jamming, eavesdropping, spoofing, etc., are common. If unaddressed, these impairments can cause serious consequences, such as connectivity loss, SNR deterioration, increased bit-error rates, reduced target detection probability, privacy and confidentiality leakage, systemic disruption, and dropped sensing accuracy.

\textit{Hardware Imperfections}: Despite great advances in design and manufacturing, hardware devices are unavoidably subject to imperfections and non-idealities that can significantly impact system performance. Aging and environmental variations are typical non-anthropogenic driving forces for hardware imperfections. Representatives include power amplifier nonlinearities, in-phase-–quadrature (I/Q) imbalances, phase noises, array calibration errors, limited resolutions of converters, clock drifts, etc. Severe issues resulting from hardware imperfections encompass constellation shift and distortion, SNR reduction, system disruption, sensing inaccuracy, etc.

\textit{Deployment and Configuration Uncertainties}: Uncertainties in deployment arise when real-world devices and components do not exactly reach the positions for which they are planned; e.g., placement errors of base stations and access points, deployment errors of sensor nodes, array calibration errors, and swarm formation errors (e.g., in highly-maneuvering unmanned aerial vehicle networks). Uncertainties in configuration arise when operating values of parameters or inputs do not exactly match their optimal values, for example, suboptimal or improper resource allocation of spectra and power. Deployment and configuration uncertainties can lead to significant and widespread outcomes, such as coverage gaps, interference, signal degradation, suboptimal or poor quality of service, inefficient resource management, and inaccurate environmental measurements (e.g., low resolution in multi-target direction-of-arrival estimation).

\textit{Timing Errors}:
Timing errors occur when synchronization between devices (e.g., between transmitters and receivers) is inaccurate. Such errors can arise from factors like clock drifts in oscillators, clock skews between different devices or systems, and time-quantization errors. If not properly addressed, timing errors can lead to severe cascading effects, for example, inter-symbol interference, inter-carrier interference (e.g., in orthogonal frequency-division multiplexing), synchronization failures in time-division multiple access (TDMA) or distributed networks, 
and poor sensing quality in time-difference-of-arrival (TDoA) and time-of-arrival (ToA) methods.

\textit{Variability of Network Topology}:
Wireless systems are inherently dynamic, with frequently changing network topologies due to, e.g., node mobility, node availability (i.e., joining and quitting), and networking protocols. The variability of network topology poses significant challenges in maintaining stable, efficient, and reliable communication connections, further leading to deficient service quality, such as message latency and dropout. Moreover, the variability can greatly impact the accuracy of networked sensing systems because, for example, the nominal topology assumed in signal processing algorithms may deviate from the underlying actual one.

\textit{Variability of Available Resources}:
Resource allocation for wireless communications and sensing focuses on optimizing the utilization of critical resources in time, space, spectrum, power, and computing. However, in real-world operations, accessible resources may be inexactly known. For example, devices relying on solar, wind, or radio energy harvesting can face power variability due to environmental factors such as weather or disasters. For another example, in spectra-sharing systems, such as cognitive radio and integrated sensing and communications (ISAC), the accessibility of spectral resources is highly variable due to, e.g., spectrum sensing inaccuracies, primary user activities, interference from coexisting functions, and environmental conditions. For the third example, insufficient computing power and memory can lead to delays and denial-of-service in executing signal processing and machine learning tasks. The variability of available resources may cause significant performance degradation in wireless systems, including decreased data throughput, increased latency, unstable connection, lower reliability (e.g., frequent system disruption), deteriorated detection and estimation accuracy, etc.

\textit{Environmental Uncertainties}:
Environmental uncertainties mean that the operating environment of a system is not completely known to model designers, algorithm developers, and decision makers, due to the inherent variability, uncontrollability, and unpredictability of environments. Environmental uncertainties act as causing factors of many other uncertainties, such as channel uncertainties, noises and interferences, variability of network topology, and variability of available resources. Rain, snow, humidity and temperature fluctuations, natural disasters, solar activities, space radiations, industrial electromagnetic interference (e.g., when machines start), vehicle moving, and human activities (e.g., gatherings, scatterings) are typical reasons for environmental uncertainties.

\textit{Algorithmic and Computational Errors}:
Algorithmic and computational errors can originate from model surrogate errors (e.g., sample-average approximation of mathematical expectation), local optimality in optimization (e.g., alternating direction method of multipliers), empirical (thus suboptimal) specifications of initial and termination conditions of iterative algorithms, numerical errors (e.g., in rounding, truncation, matrix inversion, quantization, discretization), and computational resource constraints (e.g., incomplete or delayed computations). These errors can lead to critical issues, such as unreliable predictions and decisions, inaccurate communication and sensing, system instabilities, safety risks, and insufficient management of power and spectral resources, to name a few; specific examples include localization and positioning errors, sub-optimality in resource allocation, etc. In the era of intelligent transmission and processing, as wireless systems evolve to incorporate advanced technologies like machine learning and edge computing, the risks associated with algorithmic and computational errors become progressively critical.

\textit{Data Scarcity}: The availability of sufficient and high-quality data is essential for optimal performance in tasks such as system modeling, parameter (e.g., channel) estimation, and decision-making (e.g., pattern recognition). The scarcity of data can result from inherent system limitations (e.g., sparse sensor deployments and limited sampling rates), environmental constraints (e.g., harsh or dynamic environments), privacy and security concerns (e.g., data restriction), data aging, or high cost of data acquisition. If data scarcity is not addressed, it can lead to a range of severe consequences that undermine the effectiveness of wireless systems, including inaccurate channel estimation, unreliable signal detection and estimation (e.g., in target positioning and tracking), insufficient model calibration, limited generalization ability of decisions (e.g., in data-driven machine learning), and inadequate identification and counter ability against cyberattacks.

In summary, neglecting uncertainties in wireless systems can cause significant losses of trustworthiness in terms of four critical aspects:
\begin{enumerate}
    \item \textit{Reliability}: The ability to meet performance targets despite given uncertainties. For example, high reliability can mean that the likelihood of failure is below a specified threshold. In wireless communications, minimizing the outage probability is a typical example of achieving system reliability \cite{classen2014chance}.

    \item \textit{Resilience}: The ability to return to normal states or recover expected functionality after disruptions. In wireless communications, an excellent example is as follows: a base station can hand over connections to neighboring stations when it fails or is overloaded.

    \item \textit{Adaptivity}: The ability to maintain performance by adjusting decisions, behaviors, configurations, or functionalities in response to changes in environments or conditions. In wireless communications, adaptive precoding and combining using real-time channel state information is a representative example \cite{ibnkahla2017adaptive,wang2024fast}.

    \item \textit{Robustness}: The ability to sustain performance by tolerating uncertainties; that is, the performance or output is insensitive to perturbations in parameter or input. In wireless communications and sensing, robust Capon beamforming against pointing errors and array calibration errors is a notable example \cite{vorobyov2008relationship,huang2023robust}.
\end{enumerate}
Therefore, UA wireless systems engineering seeks to ensure the trustworthiness of wireless systems, including reliability, resilience, adaptivity, and robustness.

\blue{The demonstrated types of uncertainties and their corresponding hazards commonly arise, either individually or concurrently, in diverse wireless systems, such as IoT, wireless sensor networks (WSNs), mobile ad hoc networks (MANETs), and vehicular ad hoc networks (VANETs); see \cite{ye2008robust,li2020toward,cika2020modeling}. In practice, different uncertainties are often independently addressed.}
Although the taxonomy and examples above cover a broad range of wireless engineering, they are not necessarily exhaustive. Researchers and practitioners should be aware of other types of uncertainties. 

\section{Technical Treatments Against Uncertainties}\label{sec:technical}
This section summarizes representative strategies to combat uncertainties in wireless systems. As per the characteristics of these approaches, we group them into two categories: UA designs of signals and systems, and UA modeling and computational frameworks. Some of these methods are invented from the perspective of systems structuring and design, while others are from that of systems control and operation. 

\subsection{UA Designs of Signals and Systems}\label{subsec:UA-signal-system}
\blue{In wireless communications and sensing, signal characteristics (e.g., structures, parameters), system architectures, and operational strategies play a key role in uncertainty awareness. This subsection discusses the efforts in this direction.}

\subsubsection{Redundancy, Diversity, and Margin}
Redundancy and diversity are two specific approaches of overengineering, where a system or solution is designed in a more complex and resource-consuming way than minimally required. Redundancy means adding homogeneous components or parallel systems to maintain functionality in case one fails, while diversity involves using heterogeneous components or methods to mitigate the risk of common-mode failures. In cellular networks, deploying dense homogeneous base stations and employing intelligent reflective surfaces can mitigate dead-zone and shadowing effects. In wireless sensing, applying information fusion on heterogeneous sensor grids can reduce target detection, positioning, and tracking errors under various types of interference such as jamming. Channel coding methods, e.g., Hamming codes, are excellent examples of redundancy. Temporal diversity such as interleaving, spatial diversity such as multi-input multi-output (MIMO), and frequency diversity such as orthogonal frequency-division multiplexing (OFDM) are remarkable examples of diversity. Redundancy and diversity schemes can also be jointly used. For example, time redundancy (i.e., cyclic prefixes) in OFDM further enhances the reliability of information transmission. On the other hand, margin entails introducing buffering regions to the design of signals, systems, methods, etc., to suppress uncertainties. Unlike redundancy and diversity, margin does not necessarily add complexities, although it consumes additional time, spectral, or power resources as well. When transmit power is fixed, lower-order modulation schemes, e.g., quadrature phase shift keying (QPSK), have larger I/Q margins than higher-order quadrature amplitude modulation (QAM) to combat noises and interference. Guard intervals in the time domain and guard bands in the frequency domain can add margins (i.e., buffers) to account for channel uncertainties. As illustrated, redundancy, diversity, and margin in signals and systems are natural ways to ensure reliability, resilience, and robustness. 

\subsubsection{\red{Feedback and Adaptivity}}
Adaptivity means that a system or method can monitor its running environment or conditions, learn from shifts in the environment or changes in the conditions, and adjust its structure or operation in a real-time manner. Feedback is a key component of the adaptivity loop since it informs performance or environmental changes and guides subsequent adjustments. In wireless communications, adaptive modulation and coding leveraging channel feedback, and cognitive radio for real-time spectral sensing and allocation, are two typical examples. In wireless sensing, adaptive beamforming for target tracking exemplifies the power of adaptivity and feedback; to be specific, the beamformer can steer the main lobe toward the real-time location of the target. Fast adaptation to evolving channel states (e.g., due to changing environments) is also a trending consideration in emerging machine-learning (ML) techniques for wireless communications and signal processing \cite{wang2024fast}. 

\subsubsection{Anomaly Detection and Handling}
Anomaly detection and handling is a mainstream approach to addressing uncertainties in the operation of a system, algorithm, or process. Specific examples include the following: a) error detection and correction codes in channel coding; b) error monitoring (using pilots) and reduction in signal detection, through techniques such as channel equalization and Doppler compensation (if large errors occur); c) intrusion detection and access control (e.g., intruder quarantining) in wireless networks; d) abnormal sensor detection and data imputation in sensor grids. Feedback and adaptivity represent a special case of anomaly detection and handling. Unlike feedback and adaptivity, anomaly detection and handling do not necessarily involve a feedback loop to dynamically influence the sources of errors. Instead, they may simply provide remedial solutions, such as robust methods \cite{ye2008robust,jagadeesan2018distributionally,huang2023robust}, to mitigate or tolerate these faults. 

\subsubsection{Modularization}
Modularization entails breaking down a complex system into structurally smaller, functionally independent, and operationally more manageable self-contained modules. This approach can isolate issues and failures to specific modules, so it becomes easier to locate and fix problems. In addition, replacing damaged individual parts is operationally more convenient. Moreover, modularized systems are flexible for topological reconfiguration and upgrading. Therefore, modularization can naturally benefit the improvement of the system's flexibility, scalability, and resilience. Under the modularization scheme, reliability, adaptivity, and robustness can also be enhanced in the sense that UA technical treatments for an individual module are more manageable than those for the whole system. In this sense, a highly integrated information transmission and processing (ITP) system is not necessarily preferable over its modularized counterpart; the former refers to an ML-based black-box ITP system, while the latter means a canonical communications system \cite[Figs.~2,~5]{wang2024machine}. 

\subsubsection{Decentralization}
Decentralization means distributing the computing power and data storage of a network over multiple independent nodes, rather than relying on a single central node. Decentralization can improve the network's reliability, resilience, and robustness against failures and attacks because disruptions at a subgroup of nodes may not necessarily obliterate the functionality of the entire network. Moreover, decentralization can enhance the network's scalability because joining or exiting nodes may not significantly impact the workload and strategy of a single member. In centralized networks, however, the central coordinator or controller is highly influenced by topological variability. Edge computing and federated learning are typical examples of decentralization in wireless communications. In wireless sensing, decentralization can be exemplified by distributed sensor networks. To sum up, in terms of reliability, resilience, adaptivity, and robustness, the decentralized system architecture is beneficial compared to its centralized counterpart.

\subsubsection{Prediction and Prescription}
Prediction and prescription is a strategy to forecast future events or behaviors and take proactive actions in advance to prevent or reduce adverse impacts. This strategy helps to maintain smooth functionality and reduce disruptions of a system or method, thus improving its reliability and robustness. For example, in ISAC systems, the incorporation of sensing functions enables base stations to forecast the positions of users, and therefore, generate predictive beams for smoother and robust quality of communications services. For another example, machine learning in network management can predict the traffic patterns in cellular networks (e.g., user demand spikes and traffic density in specific cells) using historical data, and therefore, enable proactive load balancing for better user experience in the future. 
For the third example, machine learning models (e.g., recurrent neural networks, long short-term memory networks) can predict future channel conditions based on historical channel state information, and therefore, empower proactive resource allocation (e.g., adaptive modulation and coding, beamforming) for smoother and robust quality of communications and sensing service in the time ahead. 

\subsection{UA Modeling and Computational Frameworks}\label{subsec:UA-frameworks}
\blue{When a wireless communication or sensing problem can be formulated as an optimization, signal processing, or machine learning model, existing analytical and computational frameworks in these domains become particularly applicable. This subsection discusses the efforts in this direction.}

\subsubsection{Uncertainty Quantification}
Let $\vec \xi$ denote a quantity of interest that takes values on real or complex coordinate spaces; for example, locations of users, channel matrices, channel noise powers, 
budgets of transmit power, and predicted future loads of base stations or access points. 
\blue{From the perspective of mathematical modeling and algorithmic computation, the initiative step of uncertainty awareness in $\vec \xi$ is to quantify uncertainties.} Most prevalent treatments in this regard include the following: a) region analysis, e.g., specifying the practically smallest region $\Xi$ on which $\vec \xi$ tasks its values; b) probability method, e.g., modeling $\vec \xi$ as a random vector or matrix that has distribution $\P_{\vec \xi}$. Note that $\Xi$ and $\P_{\vec \xi}$ can also be specific to extra conditions such as time, space, and frequency; that is, $\Xi$ and $\P_{\vec \xi}$ can be time-, position-, and frequency-selective. Other less-trending uncertainty-quantification approaches encompass fuzzy logic, Dempster--Shafer evidence theory, etc. \blue{After quantifying uncertainties, handling methods such as UA optimization, adaptive and robust signal processing, and trustworthy machine learning are the next consideration.}

\begin{table*}[!htbp]
\caption{Uncertainty-Aware Optimizations}
\label{tab:UA-opt}
\centering
\begin{tabular}{l|l|l}
\hline
\tabincell{c}{\textbf{Name}} & \tabincell{c}{\textbf{Formulation}} & \tabincell{c}{\textbf{Remarks}} \\ 
\hline
                   Original Optimization
              &       
                    $
                        \begin{array}{cl}
                           \displaystyle \min_{\vec x \in \cal X}  &  f(\vec x, \vec \xi) \\
                            \st & \vec g(\vec x, \vec \xi) \le \vec 0 
                        \end{array}
                    $
              &     \tabincell{l}{
                    $\vec x$: decision, $\cal X$: domain, $\vec \xi$: parameter, \\
                    $f$: cost function (mostly nonnegative), \\
                    $\vec g$: constraint function (possibly multi-output). 
                    }
            \\ 
\hline
                    Robust Optimization
              &       
                    $
                        \begin{array}{cl}
                           \displaystyle \min_{\vec x \in \cal X} \max_{\vec \xi \in \Xi}  &  f(\vec x, \vec \xi) \\
                            \st & \displaystyle \max_{\vec \xi \in \Xi} \vec g(\vec x, \vec \xi) \le \vec 0
                        \end{array}
                    $
              &     \tabincell{l}{
                    $\Xi$: uncertainty set of $\vec \xi$. \\
                    Philosophy: minimize \textit{worst-case} cost (i.e., avoid significant \\ performance degradation) \& guarantee \textit{worst-case} feasibility.
                    }
              \\ 
\hline
                    Stochastic Optimization
              &       
                    $
                        \begin{array}{cl}
                           \displaystyle \min_{\vec x \in \cal X}  & \E_{\vec \xi \sim \P_{\vec \xi}} f(\vec x, \vec \xi) \\
                            \st & \E_{\vec \xi \sim \P_{\vec \xi}} \vec g(\vec x, \vec \xi) \le \vec 0
                        \end{array}
                    $
              &     \tabincell{l}{
                    $\E$: expectation operator. \\
                    Philosophy: minimize \textit{expected} cost \& guarantee \textit{expected} feasibility. \\
                    \tabincell{l}{
                    Sample-Average Approx. (SAA) \\
                    Using Samples $\{\vec \xi_1, \vec \xi_2, \ldots, \vec \xi_n\}$
                    }
                    : 
                    $
                        \begin{array}{cl}
                           \displaystyle \min_{\vec x \in \cal X}  & \frac{1}{n} \sum^n_{i=1} f(\vec x, \vec \xi_i) \\
                            \st & \frac{1}{n} \sum^n_{i=1} \vec g(\vec x, \vec \xi_i) \le \vec 0.
                        \end{array}
                    $
                    }
              \\ 
\hline
                    \tabincell{l}{Chance-Constrained \\ Optimization}
              &       
                    $
                        \begin{array}{cl}
                           \displaystyle \min_{\vec x \in \cal X}  & \E_{\vec \xi \sim \P_{\vec \xi}} f(\vec x, \vec \xi) \\
                            \st & \Pr_{\vec \xi \sim \P_{\vec \xi}} [\vec g(\vec x, \vec \xi) \le \vec 0] \ge \alpha
                        \end{array}
                    $
              &     \tabincell{l}{  
                    $\Pr_{\vec \xi} [\cdot]$: probability of argument event induced by random vector $\vec \xi$. \\
                    Philosophy: require the probability of feasibility to be no less than \\ a pre-specified level $\alpha$ (e.g., $0.95$).
                    }
                \\ 
\hline
                    \tabincell{l}{Mean-Variance \\ Optimization}
              &       
                    $
                    \begin{array}{cl}
                       \displaystyle \min_{\vec x \in \cal X}  & \E_{\vec \xi \sim \P_{\vec \xi}} f(\vec x, \vec \xi) + \lambda_1 \D_{\vec \xi \sim \P_{\vec \xi}} f(\vec x, \vec \xi) \\
                        \st & \E_{\vec \xi \sim \P_{\vec \xi}} \vec g(\vec x, \vec \xi) + \lambda_2 \D_{\vec \xi \sim \P_{\vec \xi}} \vec g(\vec x, \vec \xi) \le \vec 0
                    \end{array}
                    $
              &     \tabincell{l}{  
                    $\D$: variance operator, $\lambda_1$, $\lambda_2$: trade-off parameters. \\
                    Philosophy: minimize variances for smaller variability (i.e., higher \\ robustness).
                    }
                \\ 
\hline
                    \tabincell{l}{Mean-VaR \\ Optimization}
              &       
                    $
                    \begin{array}{cl}
                       \displaystyle \min_{\vec x \in \cal X}  &  \E_{\vec \xi \sim \P_{\vec \xi}} f(\vec x, \vec \xi) + \lambda \operatorname{VaR}_\alpha[f(\vec x, \vec \xi)] \\
                        \st & \text{considerations of } \vec g(\vec x, \vec \xi) \text{ under } \P_{\vec \xi}
                    \end{array}
                    $
              &     \tabincell{l}{  
                    Value-at-Risk (VaR): 
                    $
                    \operatorname{VaR}_\alpha[h(\eta)] := \displaystyle \inf_{s:~\Pr_{\eta \sim \P_{\eta}}[h(\eta) \le s] \ge \alpha} s
                    $. \\
                    Philosophy: minimize VaR for smaller variability  (higher robustness). \\
                    Considerations of $\vec g(\vec x, \vec \xi)$ under $\P_{\vec \xi}$ can be mean, variance, VaR, \\ probability (i.e., chance-constraint), etc.
                    }
                \\ 
\hline
                    \tabincell{l}{Distributionally Robust \\ Optimization}
              &       
                    $
                        \begin{array}{cl}
                           \displaystyle \min_{\vec x \in \cal X} \max_{\P \in \cal U}  & \E_{\vec \xi \sim \P} f(\vec x, \vec \xi) \\
                            \st & \displaystyle \max_{\P \in \cal U} \E_{\vec \xi \sim \P} \vec g(\vec x, \vec \xi) \le \vec 0
                        \end{array}
                    $
              &      \tabincell{l}{
                    Philosophy: $\P_{\vec \xi}$ is not exactly known (e.g., SAA), but included in $\cal U$. \\
                    $\cal U := \{\P: d(\P, \Pb) \le \epsilon\}$: uncertainty set for $\P_{\vec \xi}$, a ball centered at $\Pb$. \\
                    $\Pb$: a reference distribution, serving as an estimate of $\P_{\vec \xi}$.
              }
              \\ 
\hline
\end{tabular}
\end{table*}

\subsubsection{UA Optimization}
Numerous communications and sensing problems can be formulated as an optimization model \cite{liu2024survey}; see \textit{Original Optimization} in Table \ref{tab:UA-opt}.  
For example, $\vec x$ can be a symbol vector in signal detection, a direction-of-arrival (DoA) vector in multi-target sensing, 
a beamformer vector in beamforming, a power control vector in resource allocation, a position vector in base-station and sensor deployment, and so on; $\vec \xi$ can denote power budgets, array steering vectors, channel matrices, constellation points, and reflection coefficients of relays or intelligent surfaces, among others.

\textit{\red{Adaptive Optimization}}: \hl{When the value of $\vec \xi$ is not exactly known, solving \textit{Original Optimization} is not practically accessible. In this case, a natural way is to estimate its real-time value and solve the problem repeatedly whenever the value of $\vec \xi$ is updated.} The estimation of $\vec \xi$ can be based on collecting more data, incorporating more expert knowledge, etc. In wireless communications, a typical example is to refine the channel state information frequently using new-coming pilots. In wireless sensing, DoA tracking is a representative instance to account for dynamic targets. In the practice of wireless engineering, the primary challenge, however, is to obtain the real-time information of $\vec \xi$ in a satisfactorily accurate manner.

\textit{Robust Optimization}: \hl{When the real-time value of $\vec \xi$ is difficult to be accurately estimated, another popular technical treatment is to study \textit{Robust Counterpart}} \cite{ye2008robust,ben2009robust,huang2023robust}; see Table \ref{tab:UA-opt}. 
To clarify, for all possible values of $\vec \xi$, the cost of decision $\vec x$ is no larger than $\max_{\vec \xi \in \Xi} f(\vec x, \vec \xi)$, and the feasibility of $\vec x$ is always ensured. Note that $f$ and $\vec g$ may only depend on part of $\vec \xi$; e.g., $f(\vec x, \vec \xi_1)$ and $\vec g(\vec x, \vec \xi_2)$ where $\vec \xi_1 \in \Xi_1$, $\vec \xi_2 \in \Xi_2$, and $\vec \xi := [\vec \xi_1; \vec \xi_2]$. \hl{Note also that adaptivity and robustness principles can be jointly used in practice \cite{huang2023robust} because adaptive estimation can be inexact as well.}

\textit{Stochastic Optimization}: Stochastic optimization, instead of considering only the domain $\Xi$, takes into account distribution $\P_{\vec \xi}$ of $\vec \xi$ and studies expectations of random quantities $f(\vec x, \vec \xi)$ and $\vec g(\vec x, \vec \xi)$; see Table \ref{tab:UA-opt}. \textit{Chance-Constrained Optimization} (CCO), in contrast, requires the probability of feasibility to be no less than a pre-specified level $0 \le \alpha \le 1$ \cite{vorobyov2008relationship,classen2014chance,kuhn2024distributionally}; see Table \ref{tab:UA-opt}. In wireless communications, studying outage probability is a particularization of CCO; see, e.g., \cite{classen2014chance}. Since $f$ and $\vec g$ may only depend on part of $\vec \xi$, the expectation operator $\E$ or the probability operator $\Pr$ can be accordingly dropped if no uncertainty is present. In addition to CCO, other variants of stochastic optimization include  \textit{Mean-Variance Optimization} and \textit{Mean-VaR Optimization} \cite{kuhn2024distributionally}; see Table \ref{tab:UA-opt}. The value-at-risk (VaR) of a random quantity $h(\eta)$, induced by a deterministic function $h$ and another random quantity $\eta$, at confidence level $\alpha$, is the lower $\alpha$-quantile of $h(\eta)$. Minimizing VaR of $f(\vec x, \vec \xi)$ implies reducing its variability (i.e., increasing robustness) because cost function $f$ is nonnegative in most wireless design and operation problems; cf. minimizing variance in mean-variance optimization. Yet another variant of stochastic optimization takes into account conditional VaR (CVaR) as a risk measure where CVaR is the mean of tail values larger than VaR of $h(\eta)$ \cite{kuhn2024distributionally}: i.e., 
$ 
    \operatorname{CVaR}_\alpha[h(\eta)] := \E\big\{h(\eta) \big| h(\eta) \ge \operatorname{VaR}_\alpha[h(\eta)]\big\}$.
Minimizing the CVaR of $f(\vec x, \vec \xi)$ can also reduce its variability, which, however, brings additional technical benefits than minimizing VaR, for example, capturing tail risk. In real-world applications, if we only have access to $n$ samples $\{\vec \xi_1, \vec \xi_2, \ldots, \vec \xi_n\}$ drawn from $\P_{\vec \xi}$, rather than the full knowledge $\P_{\vec \xi}$ itself, the stochastic-optimization counterpart can be estimated using the sample-average approximation (SAA). Stochastic-optimization variants based on other $\P_{\vec \xi}$-involved quantities, such as chance-constraint, variance, VaR, and CVaR, can be approximated using samples similarly.

\textit{Distributionally Robust Optimization}:
In stochastic optimization and its variants, we have assumed that the distribution $\P_{\vec \xi}$ of $\vec \xi$ is exactly known. In practice, however, this assumption can be highly untenable, and only a prior guess $\Pb$ of $\P_{\vec \xi}$ is available. For example, $\Pb$ can be the empirical distribution constructed using samples $\{\vec \xi_1, \vec \xi_2, \ldots, \vec \xi_n\}$. As a result, we can consider an uncertainty set $\cal U$ for distribution $\P_{\vec \xi}$ and formulate \textit{Distributionally Robust Counterpart} \cite{jagadeesan2018distributionally,kuhn2024distributionally}; see Table \ref{tab:UA-opt}. 
The uncertainty set $\cal U$ is usually constructed using a distributional ball where $d$ is a similarity measure between two distributions $\P$ and $\Pb$, and $\epsilon \ge 0$ is the radius. Intuitively, although we do not exactly know $\P_{\vec \xi}$, we assume that $\P_{\vec \xi}$ is included in $\cal U$; the more trustable the prior $\Pb$ is, that is, the closer $\Pb$ is to true distribution $\P_{\vec \xi}$, the smaller the value of $\epsilon$ should be. The distributionally robust counterpart for other $\P_{\vec \xi}$-involved quantities, such as chance-constraint, variance, VaR, CVaR, etc., can be similarly obtained, for instance, $
\min_{\P \in \cal U} \Pr_{\vec \xi \sim \P} [\vec g(\vec x, \vec \xi) \le \vec 0] \ge \alpha$.

\textit{Objective Deliberation}: As illustrated by UA optimization techniques, such as stochastic optimization and its variants, as well as (distributionally) robust optimization, the key to uncertainty awareness is to modify the objective and constraint functions. To this end, uncertainty awareness can also be pursued from the very beginning of an optimization task, that is, to deliberate the design of objective and constraint functions. For example, in regression analysis, we can replace the mean-squared-error cost function with the Huber cost function to achieve robustness against outliers \cite{zoubir2012robust}. In wireless communications and sensing, the Huber cost is particularly useful in outlier-aware signal processing and machine learning for, e.g., wireless signal estimation.

\subsubsection{\red{Adaptive and Robust Signal Processing}}
Typical roles of signal processing in wireless communications and sensing include data compression (e.g., source coding and decoding, rate-distortion analysis), modulation and demodulation, channel estimation and equalization, error detection and correction (e.g., channel coding and decoding), spectrum analysis (e.g., in cognitive radio), transmit beamforming (e.g., for resource allocation), receive beamforming and spatial filtering (e.g., for waveform, power, or DoA estimation), time synchronization, frequency alignment, noise and interference suppression (e.g., digital filtering), signal detection and estimation (e.g., in MIMO), waveform design, pattern recognition and target detection, information fusion (e.g., in sensor network), user behavior and station load forecasting, and target localization and tracking, among many others. A signal-processing process can be seen as a computational system or method with explicit input and output. As such, general UA principles in systems design can be used to improve the reliability, resilience, adaptivity, and robustness of a signal processing approach; specifically, these principles include the following: a) redundancy, diversity, and margin; b) feedback and adaptivity; c) anomaly detection and handling; d) modularization; e) decentralization; and f) prediction and prescription. When a signal processing problem is cast into an optimization formulation, UA optimization methods can also be considered; these methods encompass a) adaptive optimization, b) robust optimization, c) stochastic optimization and its variants, d) distributionally robust optimization, and e) objective deliberation. For concrete examples, see, e.g.,  \cite{ibnkahla2017adaptive,jagadeesan2018distributionally,huang2023robust,dietrich2007robust,zoubir2012robust}.

\subsubsection{\red{Trustworthy Machine Learning}}
Typical roles of machine learning in wireless communications and sensing can be largely covered by optimization and signal processing; for example, resource allocation, dynamic network management, beamforming, modulation, coding, signal detection and estimation, user behavior and station load prediction, target positioning and tracking, and channel estimation, compression, and feedback. The main difference is that machine learning heavily depends on historical observation data rather than physical mechanism modeling \cite{wang2020thirty}. The benefit of this data-driven nature is to free scientists and engineers from frustrating modeling of a complex and dynamic physical process, real-world phenomenon, or data-generating mechanism. The drawback arises, nevertheless, when the data set is scarce (e.g., due to the fast variability of wireless environments) and the computational resources are limited (e.g., in IoT devices); the less data and computational resources that we have, the less trustable a machine learning method is. Trustworthiness in machine learning for wireless communications and sensing includes, but is beyond, reliability, resilience, adaptivity, and robustness; cf. \cite[Fig.~10]{wang2024machine}. In this sense, general UA principles in systems design, optimization, and signal processing can be resorted to for trustworthy machine learning, such as redundancy, diversity, modularization, decentralization, prediction and prescription, adaptivity, robustness, objective deliberation, etc. Other trustworthiness considerations encompass explainability (e.g., transparency, fairness) and sustainability (e.g., power efficiency, ethics), which, however, require additional technical elaborations and exceed the scope of this article. 

Ad hoc technical treatments are also widely reported for trustworthy machine learning to accommodate uncertain factors; see, e.g., \cite{gawlikowski2023survey}. Typical examples include the following.
\begin{itemize}
    \item Data Engineering: to transform and augment data set. Excellent examples encompass noise injection, synonym replacement and paraphrasing of texts, rotating and flipping of images, cropping and splitting of audios, artifact interference in wireless engineering, etc.

    \item Hypothesis Engineering: to choose a suitable hypothesis space to avoid overfitting. For example, for Gaussian-distributed data, linear models are optimal and therefore sufficient. The employment of highly-nonlinear neural networks, on the contrary, tends to overfit the data. In this sense, tailoring a suitable function space (e.g., a neutral network architecture) for a specific problem is crucial.

    \item Algorithm Engineering: to opt for a suitable computational approach and control its execution. For example, adaptive learning (e.g., few-shot, transfer, continual), adversarial training, regularization, ensemble learning (e.g., bagging, boosting), dropout in neural network training, and tricks in stochastic gradient descent (e.g., momentum, gradient clipping, and early stopping) are typical choices.
\end{itemize}
The three techniques above, however, can also be incorporated into signal processing methods, e.g., \cite{pan2018noise} for noise injection.

\section{Prices of Uncertainty Awareness}\label{sec:prices}
The law of \quotemark{no free lunch} is philosophically and technically applicable to the systems engineering of almost all real-world designs, productions, and applications. Uncertainty awareness in wireless communications and sensing is no exception. Addressing uncertainties such as noise, interference, hardware imperfections, dynamic environments, data scarcity, and computational errors often requires balancing conflicting performance metrics. To be specific, optimizing for reliability, resilience, adaptability, and robustness often comes at the expense of other critical factors like computational complexity, resource efficiency, or nominal optimality. For example, some typical trade-offs can be highlighted as follows:
\begin{itemize}
    \item \textit{Robustness vs Efficiency}: Robust strategies that prioritize robustness to uncertainties, such as redundancy, diversity, and margin through employing conservative resource allocations, spectra spread, guard times, error-correction mechanisms, etc., often sacrifice resource efficiency because extra resources, which may remain idle, are occupied.

    \item \hl{\textit{Robustness vs Complexity}: By introducing overengineering techniques (e.g., redundancy and diversity) and complicated optimization formulations (e.g., Table \ref{tab:UA-opt}), the structural and computational complexities of the system, model, or method are largely increased; see \cite{ben2009robust,kuhn2024distributionally}.}

    \item \textit{Robustness vs Optimality}:
    Robustness against uncertainties loses nominal optimality under ideal or near-ideal conditions. To be specific, when significant uncertainties do not actually occur in practice since they are random, the optimality cannot be reached by a robust solution because it pursues optimality under the worst case, not under the nominal situation.

    \item \textit{Modularization vs Integration}: To isolate faults and facilitate maintenance, we prefer the modularization scheme. Yet, for the purposes of resource sharing and overall optimality, e.g., in ISAC and ML-based ITP systems, we may resort to integration.

    \item \textit{Centralization vs Decentralization}: 
    Centralized systems enable efficient global optimality of resource use and node interaction, but are prone to central-node failure and network attacks. Decentralized systems enhance robustness and scalability by distributing data storage and computation, but may face node-coordination challenges and suboptimal global performance.
\end{itemize}

\section{\blue{Examples of Uncertainty Handling}}\label{sec:examples}

\blue{In Subsection \ref{subsec:UA-signal-system}, concrete examples of UA designs of signals and systems are provided. This section presents additional examples that leverage UA modeling and computational frameworks in Subsection \ref{subsec:UA-frameworks}.}

\textit{\blue{Fast Adaptations for Wireless Communications}}:
In wireless communications, channel conditions vary over time due to environmental fluctuations and user mobility. Consequently, the entire wireless transmission process must adapt to real-time channel conditions, including channel estimation, transmit precoding, receive combining, and resource allocation, among others \cite{ibnkahla2017adaptive,wang2024fast}. Two immediate challenges arise: first, acquiring sufficient new pilot data to estimate updated channel state information; second, ensuring adequate computing power for adaptive operations. However, in practice, both pilot data and computing resources are limited, especially for highly dynamic channels and resource-constrained edge devices. Hence, computationally-lightweight and data-efficient methods in wireless transmission must be developed to handle channel evolution. In signal processing for wireless communications, typical examples include adaptive modulation and coding, adaptive beamforming, and adaptive receivers, among others \cite{ibnkahla2017adaptive}. In machine learning for wireless communications, few-shot learning has emerged as a promising approach. By leveraging meta-learning techniques, multi-task learning, and domain knowledge, few-shot learning enables computation- and data-efficient wireless solutions such as channel estimation, precoding, and signal detection \cite{wang2024fast}. 

\textit{\blue{Robust Adaptive Beamforming}}: Adaptive beamforming plays a key role in wireless communications and sensing to enhance signal-to-interference-plus-noise ratios (SINRs); Capon beamforming is a typical approach in this domain, which is formulated as $\min_{\vec w} \vec w^\H \math R \vec w,~\st~\vech a^\H \vec w = 1$, where $\math R$ denotes the estimated covariance matrix of array snapshots, $\vec w$ a beamformer, and $\vech a$ the assumed steering vector of the signal of interest \cite{huang2023robust}. In practice, $\math R$ and $\vech a$ can be inaccurate. As such, robust Capon beamforming is formulated as
\begin{equation}\label{eq:robust-capon}
    \begin{array}{cl}
       \displaystyle \min_{\vec w}  \max_{\mat R \in \cal U_R}&  \vec w^\H \mat R \vec w \\
         & \displaystyle \min_{\vec a \in \cal U_a}  \vec w^\H \vec a \vec a^\H \vec w  \ge 1,
    \end{array}
\end{equation}
to optimize the worst-case performance and ensure worst-case feasibility \cite{huang2023robust}, where $\cal U_R$ and $\cal U_a$ are the uncertainty sets (i.e., uncertainty quantifications) of $\mat R$ and $\vec a$, respectively; for example, $\cal U_R \defeq \{\mat R:~\|\mat R - \math R\|_F \le \epsilon_1\}$ and $\cal U_a \defeq \{\vec a:~\|\vec a - \vech a\|_2 \le \epsilon_2\}$
for some uncertainty levels $\epsilon_1, \epsilon_2 \ge 0
$ where $\|\cdot\|_F$ and $\|\cdot\|_2$ denote the matrix Frobenius norm and the vector $2$-norm, respectively. A chance-constraint alternative is proposed in \cite{vorobyov2008relationship}, where $\Pr_{\vec a \sim \P_{\vec a}}[\vec w^\H \vec a \vec a^\H \vec w  \ge 1] \ge \alpha$ is employed with an uncertainty-quantification distribution $\P_{\vec a}$ and a pre-specified level $\alpha$. It is reported that such UA formulations can greatly enhance beamforming performance when uncertainties are present in the estimated snapshot covariance $\math R$ and assumed steering vector $\vech a$ \cite{vorobyov2008relationship,huang2023robust}. However, this benefit comes with extra prices, such as computational burden, because solving \eqref{eq:robust-capon} requires additional efforts \cite{huang2023robust}. \hl{To sum up, through this example, the concepts of uncertainty quantification, robust optimization, chance-constrained optimization, and robustness-complexity trade-off have been illustrated.}

\textit{\blue{Distributionally Robust Localization}}: Wireless localization is a fundamental technique in IoT and WSNs. A wireless localization problem can be formulated as $\min_{\vec r} \E_{\vec r_0 \sim \Ph_{\vec r_0}} [g(\vec r, \vec r_0)]$, where $\vec r$ is an estimated location of a target, $\vec r_0$ is the actual but unknown location of this target, $g$ is a cost function such as the mean-squared error, and $\Ph_{\vec r_0}$ is an assumed uncertainty-quantification distribution of $\vec r_0$ \cite{jagadeesan2018distributionally}. In practice, the assumed distribution $\Ph_{\vec r_0}$ can be inaccurate. As such, the distributionally robust localization problem can be formulated as
\begin{equation}\label{eq:localization}
    \min_{\vec r} \max_{\P_{\vec r_0} \in \cal U} \E_{\vec r_0 \sim \P_{\vec r_0}} [g(\vec r, \vec r_0)],
\end{equation}
where $\cal U := \{\P_{\vec r_0}:d(\P_{\vec r_0}, \Ph_{\vec r_0}) \le \epsilon\}$ is the uncertainty set (i.e., uncertainty quantification) of $\P_{\vec r_0}$, $d$ is a metric between distributions, and $\epsilon \ge 0$ is the uncertainty level. It is reported that such a UA formulation can significantly reduce localization errors when distributional uncertainties are present in the assumed distribution $\Ph_{\vec r_0}$ \cite{jagadeesan2018distributionally}. However, if the assumed $\Ph_{\vec r_0}$ is accurate, solving \eqref{eq:localization} may sacrifice the performance optimality because the solution to the robust counterpart \eqref{eq:localization} does not necessarily solve the original problem $\min_{\vec r} \E_{\vec r_0 \sim \Ph_{\vec r_0}} [g(\vec r, \vec r_0)]$, leading to a conflict between robustness and optimality \cite[Fig.~5]{jagadeesan2018distributionally}. \hl{In summary, through this example, the concepts of uncertainty quantification, distributionally robust optimization, and robustness-optimality trade-off have been illustrated.}

\blue{In real-world systems such as IoT, WSNs, MANETs, and VANETs, the demonstrated uncertainties and countermeasures are applicable because signal detection, precoding and beamforming, and localization are key enabling techniques.}

\section{Conclusions}\label{sec:conclusion}
This article focuses on uncertainty awareness in wireless communications and sensing. First, the sources and complications of uncertainties are identified, for example, the reliability, resilience, adaptivity, and robustness issues resulting from channel uncertainties, interferences, cyberattacks, hardware imperfections, dynamic environments, data scarcity, and computational errors. Second, existing and developing UA
technical treatments are highlighted, including UA designs of signals and systems (i.e., redundancy, diversity, and margin, feedback and adaptivity, anomaly detection and handling, modularization, decentralization, and prediction and prescription), as well as UA modeling and computational frameworks (i.e., uncertainty quantification, UA optimization, adaptive and robust signal processing, and
trustworthy machine learning). Third, trade-offs that balance conflicting performance metrics in using UA strategies are exemplified, encompassing robustness vs efficiency, robustness vs complexity, robustness vs optimality, modularization vs integration, and centralization vs decentralization, among others. 
\blue{Fourth, motivating examples of how to apply these UA strategies are provided.}  
Through the above efforts, we deliver the following message to the community: uncertainty awareness is an enabling and pivotal aspect of intelligent transmission and processing.

\bibliographystyle{IEEEtran}
\bibliography{References}

\end{document}